\documentclass[onecolumn,showpacs,amsfonts,aps,prc,nofootinbib,floatfix,%
superscriptaddress]{revtex4}
\usepackage{epsfig}
\usepackage{amsmath}
\usepackage{bm}
\usepackage{graphicx}

\newcommand{\beq}{\begin{equation}}
\newcommand{\eeq}{\end{equation}}
\newcommand{\bea}{\vspace{0.25cm}\begin{eqnarray}}
\newcommand{\eea}{\end{eqnarray}}

\newcommand{\pb}{{{\bf p}}}

\def\lsim{\mathrel{\rlap{\lower4pt\hbox{\hskip1pt$\sim$}}
    \raise1pt\hbox{$<$}}}         
\def\gsim{\mathrel{\rlap{\lower4pt\hbox{\hskip1pt$\sim$}}
    \raise1pt\hbox{$>$}}}         

\long\def\symbolfootnote[#1]#2{\begingroup%
\def\thefootnote{\fnsymbol{footnote}}\footnotemark[#1]\footnotetext[#1]{#2}\endgroup}

\newcommand{\landau}{L.D.~Landau Institute for Theoretical Physics,
        GSP-1, 117940, Kosygina Str. 2, 117334 Moscow, Russia}

\begin{document}

\title{
Jet quenching in $pp$ and $pA$ collisions
\symbolfootnote[1]{Talk at XIth Quark
Confinement and the Hadron Spectrum, Saint-Petersburg, Russia, 8-12 September 2014.}
}


\author{B.G.~Zakharov}\affiliation{\landau}

\begin{abstract}
We study jet quenching in $pp$ and $pA$ collisions in the scenario with
formation of a mini quark-gluon plasma. 
We find a significant suppression effect.  For light hadrons
at $p_{T}\sim 10$ GeV we obtained the reduction of the spectra
by $\sim [20-30,25-35,30-40]$\% in $pp$ collisions at 
$\sqrt{s}=[0.2, 2.76,7]$ TeV. 
We discuss how  jet quenching in $pp$ collisions may change the 
predictions for the nuclear modification
factors in $AA$ collisions for light and heavy flavors.
We also give predictions for modification of the photon-tagged 
and inclusive jet fragmentation functions in high multiplicity $pp$ 
events. 

\end{abstract}


\maketitle

\section{Introduction}
One of the manifestation of the quark-gluon plasma (QGP) formation in
$AA$ collisions is the jet quenching phenomenon which
is dominated by the radiative parton energy loss
\cite{GW,BDMPS,LCPI,BSZ,W1,GLV1,AMY}.
It leads to suppression of the high-$p_{T}$ spectra, which 
is characterized by the nuclear
modification factor $R_{AA}$ given by the ratio
of the inclusive cross section for $AA$ collisions to the binary-scaled
inclusive cross section for $pp$ collisions
\beq
  R_{AA} = \frac{d\sigma(AA\to hX)/d\pb_{T}dy}{N_{bin} \, 
  d\sigma(pp\to hX)/d\pb_{T}dy}\,.
\label{eq:10}
\eeq
It would be extremely interesting to observe jet quenching in 
$pp$ and $pA$ collisions, since it would be a direct signal of the mini-QGP
formation.
The QGP formation in $pp$ and $pA$ collisions 
have been addressed in several
publications recently \cite{Bozek_pp,Wied_pp,glasma_pp} from the viewpoint
of the hydrodynamical flow effects.
In recent papers \cite{Z_phot,Z_RPP} we studied the possible manifestations
of jet quenching in $pp$ collisions within 
the light-cone path integral approach \cite{LCPI}, which we previously
used for analysis of jet quenching in $AA$ collisions 
\cite{RAA08,RAA11,RAA12,RAA13}. In \cite{Z_phot} we discussed
the medium modification of the $\gamma$-tagged fragmentation functions (FFs)
and in \cite{Z_RPP} the medium modification factor $R_{pp}$ and its
effect on the nuclear modification factors $R_{AA}$ and $R_{pA}$. 
The medium modification factor $R_{pp}$ characterizes the difference
between the real inclusive $pp$ cross section, 
accounting for the final-state jet interaction in the QGP, 
and the perturbative one, i.e.,
\beq
d\sigma(pp\to hX)/d\pb_{T}dy=
R_{pp} d\sigma_{pert}(pp\to hX)/d\pb_{T}dy\,.
\label{eq:20}
\eeq
Since we cannot switch off the final state interaction in the QGP, 
the $R_{pp}$ is not an observable quantity. Nevertheless, it may affect
the theoretical predictions for $R_{AA}$.
Indeed, in the scenario with the QGP formation in $pp$ collisions 
one should use in the denominator in (\ref{eq:10}) 
the real inclusive $pp$ cross section which differs from the perturbative
one. In this case one should compare with experimental
$R_{AA}$ the following quantity:
\beq
R_{AA}=R_{AA}^{st}/R_{pp}\,,
\label{eq:21}
\eeq
where $R_{AA}^{st}$ is the standard  nuclear modification factor 
calculated using the pQCD predictions for the particle spectrum in 
$pp$ collisions.
The effect of the $R_{pp}$ may be 
important for the centrality dependence of $R_{AA}$
and the azimuthal anisotropy 
(simply because in the scenario with the QGP formation 
in $pp$ collisions $\alpha_{s}$ becomes bigger).
It should also be important for the jet flavor 
tomography of the QGP \cite{Armesto_HQ,BG,RAA12,RAA13}.
Because the effect of $R_{pp}$ on
$R_{AA}$ for heavy quarks should be smaller due to weaker jet quenching
for heavy quarks in $pp$ collisions.
In this talk I review the results of \cite{Z_phot,Z_RPP} and  
extend the analysis \cite{Z_RPP} to heavy flavors.

\section{mini-QGP in proton-proton collisions}
We describe the mini-QGP fireball 
within 1+1D Bjorken's model \cite{Bjorken2}, which gives
$T_{0}^{3}\tau_{0}=T^{3}\tau$. 
For $\tau<\tau_{0}$ we assume that the medium density $\propto \tau$. 
As in our previous analyses of jet quenching in $AA$ collisions 
\cite{RAA08,RAA11,RAA12,RAA13},
in the basic variant we take $\tau_{0}=0.5$ fm. 
For the QGP in $AA$ collisions with the lifetime/size $L\gg \tau_{0}$ 
the medium effects are not very sensitive to 
variation of $\tau_{0}$. But this may be untrue for $pp$ collisions when 
the plasma size is considerably smaller.
To understand the sensitivity
of $R_{pp}$ to $\tau_{0}$
we also perform calculations for $\tau_{0}=0.8$ fm.
To simplify the computations we neglect 
variation of the initial temperature $T_{0}$ with the 
transverse coordinates.
We fix $T_{0}$ using the entropy/multiplicity ratio 
$C=dS/dy{\Big/}dN_{ch}/d\eta\approx 7.67$ obtained in \cite{BM-entropy}.
The initial entropy density can be written as 
\beq
s_{0}=\frac{C}{\tau_{0}\pi R_{f}^{2}}\frac{dN_{ch}}{d\eta}\,,
\label{eq:30}
\eeq
where $R_{f}$ is the fireball radius.
We ignore the azimuthal anisotropy, and regard $R_{f}$ as an effective
mini-QGP radius, which includes $pp$ collisions in the whole range
of the impact parameter.
This approximation seems to be plausible since the jet production 
should be dominated by the nearly head-on collisions
for which the azimuthal effects should be weak.

In jet quenching calculations for the multiplicity density in 
(\ref{eq:30}) one should use the multiplicity density of the soft 
(underlying-event (UE)) hadrons, which is bigger than the minimum bias
multiplicity density by a factor ($K_{ue}$) of $\sim 2$ \cite{CDF}.
Experimental studies \cite{CDF,PHENIX_dA,ATLAS_UE_Nch,CMS_UE_Nch,ALICE_UE_Nch} 
show that the UE multiplicity grows with momentum of the 
leading charged jet hadron at $p_{T}\lsim 3-5$ GeV and then flattens out. 
The plateau region corresponds approximately to 
$E_{jet}\gsim 15-20$ GeV.
To fix the $dN_{ch}/d\eta$ in (\ref{eq:30}) at $\sqrt{s}=0.2$ TeV 
we use the UE enhancement factor $K_{ue}$ 
from PHENIX \cite{PHENIX_dA} obtained by dihadron correlation
method. Taking for minimum bias 
non-diffractive events $dN_{ch}^{mb}/d\eta=2.98\pm 0.34$ from STAR 
data \cite{STAR-dnch}, we obtained for the UEs in the plateau region 
$dN_{ch}/d\eta\approx 6.5$.  
To evaluate the UE multiplicity  
at $\sqrt{s}=2.76$ and $5.02$ TeV we use the data from ATLAS 
\cite{ATLAS_UE_Nch} at $\sqrt{s}=0.9$ and $7$ TeV that give in the 
plateau region $dN_{ch}/d\eta\approx 7.5$ and $13.9$.
Assuming that $dN_{ch}/d\eta\propto s^{\delta}$, by interpolating
between $\sqrt{s}=0.9$ TeV and 7 TeV we obtained for  
the UE multiplicity density in the plateau region $dN_{ch}/d\eta\approx 10.5$
and $12.6$ at $\sqrt{s}=2.76$ and $5.02$ TeV, respectively.
We use for $R_{f}$ the values obtained 
in numerical simulations of $pp$ collisions at $\sqrt{s}=7$ TeV 
performed in \cite{glasma_pp} 
within the IP-Glasma model \cite{IPG12}. 
In \cite{glasma_pp} it has been found that $R_{f}$
grows approximately as linear function of $(dN_{g}/dy)^{1/3}$ and then 
flattens out (a convenient parametrization of $R_{f}$ from \cite{glasma_pp}
has been given in \cite{RPP}). 
The plateau region corresponds to nearly head-on
collisions where the fluctuations of multiplicity are 
dominated by the fluctuations of the glasma color fields \cite{glasma_pp}. 
With the help of the formula for $R_{f}$ from \cite{RPP} for 
the above values of the UE multiplicity densities 
in the plateau regions we obtain (we take $dN_{g}/dy=\kappa dN_{ch}/d\eta$
with $\kappa=C45/2\pi^{4}\xi(3)\approx 2.13$) 
\beq
R_{f}[\sqrt{s}=0.2,2.76,5.02,7\,\, \mbox{TeV}]
\approx[1.3,1.44,1.49,1.51]\,\,\mbox{fm}\,.
\label{eq:40}
\eeq
We neglect possible variation of the $R_{f}$ from 
RHIC to LHC since our results are not very sensitive to $R_{f}$.
Using (\ref{eq:30}) and the ideal gas 
formula $s=(32/45+7N_{f}/15)T^{3}$ (with $N_{f}=2.5$),
we obtain the initial temperatures of the QGP
\beq
T_{0}[\sqrt{s}=0.2,2.76,5.02,7\,\,\mbox{TeV}]
\approx[199,217,226,232]\,\,\mbox{MeV}\,.
\label{eq:50}
\eeq
One sees that the values of $T_{0}$ lie well above 
the deconfinement temperature $T_{c}\approx 160-170$ MeV. 

For initial temperatures (\ref{eq:50}) the purely plasma phase may exist up to 
$\tau_{QGP}\sim 1-1.5$ fm. At $\tau>\tau_{QGP}$ the hot QCD matter 
will evolve in the mixed phase up to $\tau_{max}\sim 2R_{f}$ 
where the transverse
expansion should lead to fast cooling of the fireball. For 
$\tau_{QGP}< \tau < \tau_{max}$ the QGP fraction 
in the mixed phase is approximately $\propto 1/\tau$ \cite{Bjorken2}, 
and for this reason we can use $1/\tau$ dependence of the number
density of the scattering centers in the whole range of $\tau$
(but with the Debye mass defined for $T\approx T_{c}$ at $\tau> \tau_{QGP}$).

The central question for the scenario with mini-QGP formation is 
the extend to which the mini-fireball created in $pp$ collisions 
may be treated as a continuous macroscopic medium.
This question at present is still open. 
The lattice studies support 
the idea that a collective medium may be created in $pp$ collisions.
Indeed, the macroscopic behavior of the fireball is possible
when the Knudsen number $Kn\sim \tau_c/\tau$ is small. We estimated $Kn$ 
using the recent lattice results \cite{sigma_Amato} on the electric 
conductivity $\sigma$ of the QGP.
From the Drude formula (for massless partons)
\beq
\sigma\sim \frac{\langle e^{2}_{q}\rangle n_{q+\bar{q}}\tau_c}{3T}
\label{eq:51}
\eeq 
and lattice $\sigma$ from \cite{sigma_Amato}
we obtained approximately for the temperatures given in (\ref{eq:50})
$Kn(\mbox{quark})\sim 1$ 
at $\tau\sim 0.5$ fm
and $Kn(\mbox{quark})\sim 0.25$ at $\tau\sim 1$ fm. The gluon Knudsen
number should be smaller by a factor of $\sim C_{F}/C_{A}=4/9$. This
qualitative analysis shows that the collective behavior of the mini-fireball
does not seem to be unrealistic. Of course, the inequality $Kn\ll 1$ is just
a necessary condition for the hydrodynamic behavior of the QGP. But it cannot 
guarantee that the QGP is produced quickly after $pp$ collision.

\section{Medium induced gluon spectrum and parameters of the model}
As in \cite{RAA08}, we evaluate the medium induced gluon spectrum
$dP/dx$ ($x=\omega/E$ is the gluon fractional momentum)
for the QGP modeled by a 
system of the static Debye  screened color centers \cite{GW}. 
We use the Debye mass obtained in the lattice analysis \cite{Bielefeld_Md} 
giving $\mu_{D}/T$ slowly decreasing with $T$  
($\mu_{D}/T\approx 3.2$ at $T\sim T_{c}$, $\mu_{D}/T\approx 2.4$ at 
$T\sim 4T_{c}$). 
For the plasma quasiparticle masses of light quarks and gluon
we take $m_{q}=300$ and $m_{g}=400$ MeV  supported by 
the analysis of the lattice data \cite{LH}.
Our results are
not very sensitive to $m_{g}$, and practically
insensitive to the value of $m_{q}$.
For gluon emission from a quark (or gluon) the $x$-spectrum may be written
\cite{Z04_RAA}
through the light-cone wave function 
of the $gq\bar{q}$ (or $ggg$) system in the coordinate $\rho$-representation.
Its $z$-dependence is governed by
a two-dimensional Schr\"odinger equation 
with the ``mass'' $\mu=x(1-x)E$ ($E$ is the initial parton energy)
in which the
longitudinal coordinate $z$ 
plays the role of time and the potential $v(\rho)$ is proportional 
to the QGP density/entropy times a linear combination of
the dipole cross sections $\sigma(\rho)$, $\sigma((1-x)\rho)$
and $\sigma(x\rho)$.
We perform calculations 
with running $\alpha_{s}$ frozen at some value $\alpha_{s}^{fr}$ at low momenta.
For gluon emission in vacuum a reasonable choice is 
$\alpha_{s}^{fr}\sim  0.7-0.8$ \cite{NZ_HERA,DKT}. 
In plasma thermal effects can suppress $\alpha_{s}^{fr}$.
However, the uncertainties of jet quenching calculations are large and
the extrapolation from the vacuum gluon 
emission  to the induced radiation may be unreliable.
 For this
reason we treat $\alpha_{s}^{fr}$ as a free parameter
of the model. 
In \cite{RAA13} we have observed  that data on $R_{AA}$ 
are consistent with $\alpha_{s}^{fr}\sim 0.5$ for RHIC
and $\alpha_{s}^{fr}\sim 0.4$ for LHC.
The reduction of $\alpha_{s}^{fr}$ from RHIC to LHC 
may be due to stronger thermal effects at LHC where 
the initial temperature is bigger. 
But the analysis \cite{RAA13} is performed ignoring the 
medium suppression in $pp$ collisions.
Accounting for $R_{pp}$ should increase $\alpha_{s}^{fr}$.
However, in \cite{RAA13} we used the plasma density 
vanishing at $\tau<\tau_{0}$, whereas now we use 
the QGP density $\propto \tau$, which leads
to somewhat stronger medium suppression.
As a result, preferable $\alpha_{s}^{fr}$ (from the standpoint of the 
description of $R_{AA}$)
remains approximately the same, or a bit larger, as obtained 
in \cite{RAA13}.
If the difference between 
$\alpha_{s}^{fr}$  for $AA$ collisions at RHIC and LHC
is really due to the thermal effects, 
then  for the mini-QGP with $T_{0}$ as given in 
(\ref{eq:50}) a reasonable window  is 
$\alpha_{s}^{fr}\sim 0.6-0.7$. 
In principle for the mini-QGP the thermal reduction 
of $\alpha_{s}$ may be smaller than for the large-size 
plasma (at the same temperature). 
Because for the mini-QGP a considerable contribution to the induced gluon 
emission comes from the product of the emission amplitude 
and complex conjugate one when one of them has the gluon emission 
vertex outside the medium and is not affected by the medium effects.
We perform the calculations for
$\alpha_{s}^{fr}=0.5$, $0.6$ and $0.7$.
Note that $R_{pp}$ should be less sensitive to
$\alpha_{s}^{fr}$ than $R_{AA}$ since the typical
virtualities for induced gluon emission in the mini-QGP are
larger than that in the large-size QGP (see below).

The physical pattern of induced gluon emission in the mini-QGP
differs somewhat from that for the large-size QGP.
For the mini-QGP when the typical path length in the medium 
$L\sim 1-1.5$ fm the energy loss is dominated by gluons with
$L_{f}\gsim L$, where $L_{f}\sim 2\omega/m_{g}^{2}$ is the gluon formation
length in the low density limit. 
In this regime the dominating contribution 
comes from the $N=1$ rescattering,
and the finite-size and Coulomb effects play a crucial role 
\cite{Z_OA,AZ} (see also \cite{Arnold_OA}).
On the contrary, for the QGP in $AA$ collisions 
the induced energy loss is dominated by gluons with $L_{f}\lsim L$.
Indeed, 
$L_{f}\sim 2\omega S_{LPM}/m_{g}^{2}$,
where $S_{LPM}$ is the LPM suppression factor.
For RHIC and LHC typically $S_{LPM}\sim 0.3-0.5$ for $\omega\sim 2$ GeV, it 
gives $L_{f}\sim 1.5-2.5$ fm which is smaller than the typical $L$
for the QGP in $AA$ collisions.
In this regime the finite-size effects are much less important
and the gluon spectrum is (locally) approximately similar to that 
in an infinite extent matter.
It is important that 
the induced gluon emission in the mini-QGP
is more perturbative than in the large-size QGP.
Indeed, 
from the Schr\"odinger diffusion relation one can obtain for 
the typical transverse size of the three parton system 
$\rho^{2}\sim 2\xi/\omega$, where 
$\xi$ is the path length after gluon emission.
Then, using the fact that $\sigma(\rho)$ is dominated by the $t$-channel
gluon exchanges with virtualities up to $Q^{2}\sim 10/\rho^{2}$  
\cite{NZ_piki} we obtain $Q^{2}\sim 5\omega/\xi$. For $\omega\sim 2$
and $\xi\sim 0.5-1$ fm it gives $Q^{2}\sim 2-4$
GeV$^{2}$. The virtuality scale in the gluon emission
vertex has a similar form but smaller by a factor of $\sim 2.5$ \cite{Z_Ecoll}.
The $1/\xi$ dependence of $Q^{2}$ persists up to $\xi\sim L_{f}$.
For the large-size QGP one should replace $\xi$
by the real in-medium $L_{f}$ (which contains $S_{LPM}$) which
is by a factor of $\sim 2$ larger than the typical values of $\xi$ 
for the mini-QGP. It results in a factor of $\sim 2$
smaller virtualities in $AA$ collisions.

\section{Energy loss in the mini-QGP}
In Fig.~1 we show the energy dependence of the total 
(radiative plus collisional) and collisional
energy loss for partons produced in the center of the mini-QGP 
fireball for $\alpha_{s}^{fr}=0.6$ (as in \cite{Z_Ecoll}, both the radiative 
and collisional 
contributions are defined for the lost energy smaller than half of 
the initial parton energy). 
We present results for the fireball parameters obtained for the 
jet energy dependent UE $dN_{ch}/d\eta$ and for that in the plateau region
(details see in \cite{Z_RPP}). 
One can see that
the energy loss for these two versions (solid and long-dashed lines) 
become very close to each
other at $E\gsim 10$ GeV. 
Our results show that at $E\sim 10-20$ GeV for gluons the total energy loss 
is $\sim 10-15$\% of the initial energy.
The contribution of the collisional mechanism is relatively small. 
The energy loss for the mini-QGP is smaller than
that for the large-size QGP in $AA$ collisions  obtained in \cite{RAA13} 
by a factor of $\sim 4$. 

%
\begin{figure}[h]
\epsfig{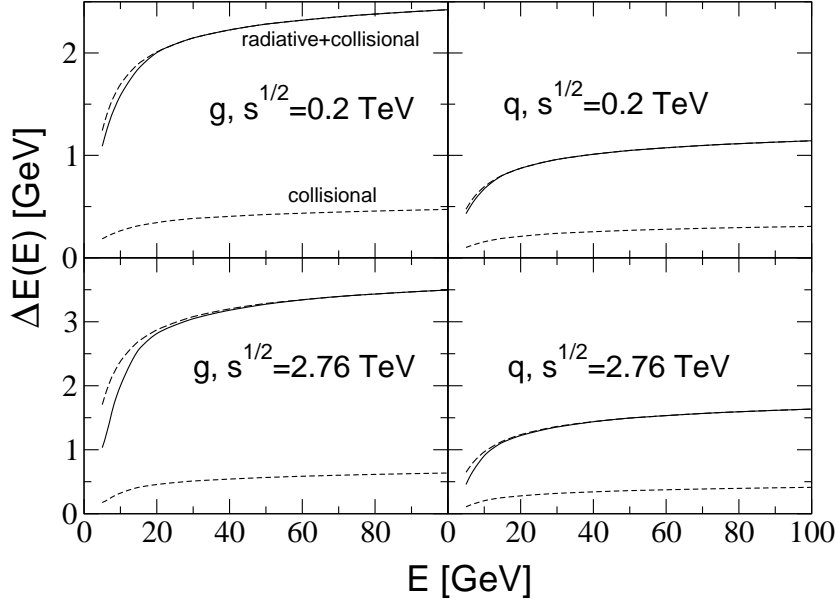}
\caption{Energy dependence of the  
energy loss of gluons (left) and light quarks (right) 
 produced in the center of the mini-QGP 
fireball at $\sqrt{s}=0.2$ TeV (upper panels) and $\sqrt{s}=2.76$ TeV (lower
panels). Solid line: total (radiative plus collisional) energy loss 
calculated  with the fireball radius $R_{f}$ and the initial temperature $T_{0}$
obtained with the UE $dN_{ch}/d\eta$ dependent on the initial parton
energy $E$; dashed line: same as solid line but for collisional energy loss;
 long-dashed line: same as solid line but for $R_{f}$ and $T_{0}$ 
obtained with the
UE $dN_{ch}/d\eta$ in the plateau region as given by (\ref{eq:40})
and (\ref{eq:50}).
All the curves are for $\alpha_{s}^{fr}=0.6$. 
}
\end{figure}
\begin{figure*}[h]
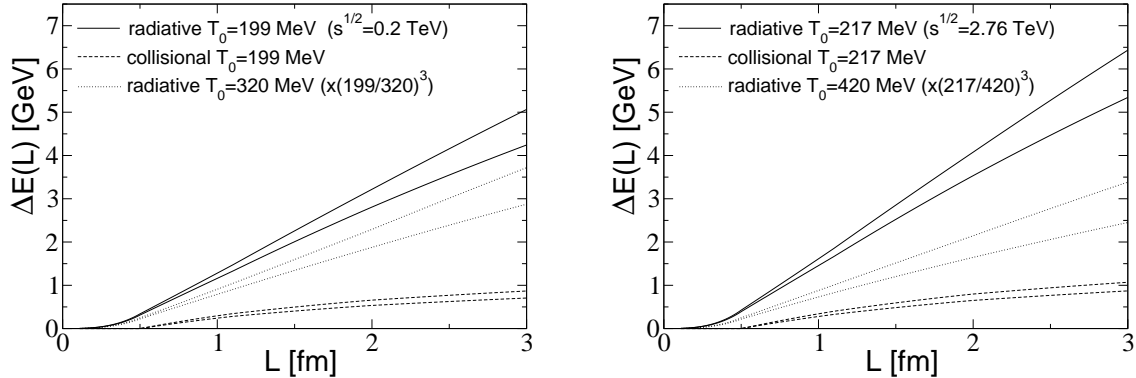

\hspace*{-0.8cm }\epsfig{file=fig2na-t.eps,height=5cm,clip=,angle=0} 
\hspace*{0.8cm } \epsfig{file=fig2nb-t.eps,height=5cm,clip=,angle=0} 
\begin{minipage}[t]{17.cm}
\caption{
Left:
Radiative (solid) and collisional (dashed) 
gluon energy loss   vs the path length $L$ in the 
QGP with $T_{0}=199$ MeV for (bottom to top)  $E=20$ and $50$ GeV.
The dotted lines show radiative energy loss for $T_{0}=320$ MeV
rescaled by the factor $(199/320)^{3}$. 
All curves are calculated for $\alpha_{s}^{fr}=0.6$.
Right: same as in the left figure but for $T_{0}=217$ and $420$ MeV
and the rescaling factor $(217/420)^{3}$ for dotted lines. 
}
\end{minipage}
\end{figure*}

In Fig.~2 we show the the radiative and collisional gluon energy loss vs 
the path length $L$
for $E=20$ and $50$ GeV
for $T_{0}=199$ and $217$ MeV, corresponding to $\sqrt{s}=0.2$ 
and $2.76$ TeV.
To illustrate the difference between  $pp$ and $AA$ collisions 
we present also predictions for radiative energy loss 
for $T_{0}=320$ MeV corresponding to central $Au+Au$ collisions
at $\sqrt{s}=0.2$ TeV, and
for $T_{0}=420$ MeV corresponding to central $Pb+Pb$ collisions
at $\sqrt{s}=2.76$ TeV. 
We rescaled the predictions for $AA$ collisions by 
the factor $(T_{0}(pp)/T_{0}(AA))^{3}$.
One sees that at $L\ge \tau_{0}$ the radiative energy loss
is approximately a linear function of $L$, and  at $L<\tau_{0}$
the radiative energy loss is approximately $\propto L^{3}$
(since the leading $N=1$ rescattering term to the effective
Bethe-Heitler cross section is $\propto L$ \cite{Z_OA,AZ} and integration 
over the longitudinal coordinate of the scattering center gives additional 
two powers of $L$).
From comparison of the radiative energy loss for $T_{0}=199$ and $217$ 
MeV to that for $T_{0}=320$ and $420$ MeV one can see a deviation from 
the $T^{3}$  scaling by factors of $\sim 1.5$ and $\sim 2$, respectively. 
This difference persists even 
at $L\sim 1$ fm.
It comes mostly from
the increase of the LPM suppression (and partly from the increase
of the Debye mass) for the QGP produced in $AA$ collisions.

\section{Medium modification of the inclusive spectra}
\subsection{Perturbative and medium modified inclusive cross sections}
As usual we write the perturbative inclusive cross section in (\ref{eq:20}) 
in terms of the vacuum parton$\to$hadron FF $D_{h/i}$
\beq
\frac{d\sigma_{pert}(pp\to hX)
}{d\pb_{T} dy}=
\sum_{i}\int_{0}^{1} \frac{dz}{z^{2}}
D_{h/i}^{}(z, Q)
\frac{d\sigma(pp\to iX)}{d\pb_{T}^{i} dy}\,,
\label{eq:60}
\eeq
where
${d\sigma(pp\to iX)}/{d\pb_{T}^{i} dy}$ 
is the ordinary hard cross section,
$\pb_{T}^{i}=\pb_{T}/z$ is the parton 
transverse momentum. 
We write the real inclusive cross section 
in a similar form but with the medium modified FF
$D_{h/i}^{m}$
\beq
\frac{d\sigma_{}(pp\rightarrow hX)}{d\pb_{T} dy}=
\sum_{i}\int_{0}^{1} \frac{dz}{z^{2}}
D_{h/i}^{m}(z, Q)
\frac{d\sigma(pp\rightarrow iX)}{d\pb_{T}^{i} dy}\,.
\label{eq:70}
\eeq
Here it is implicit that $D_{h/i}^{m}$ 
is averaged over the geometry of the parton process and over 
the impact parameter of  $pp$ collision.
%


We calculated the hard cross sections 
in  the LO pQCD with the CTEQ6 \cite{CTEQ6} parton distribution functions 
(PDFs).
To simulate the higher order effects we calculate the partonic
cross sections for the virtuality scale of $\alpha_{s}$  
$cQ$ with $c=0.265$ as in the PYTHIA event generator \cite{PYTHIA}.
For the hard scale $Q$ in the FFs in (\ref{eq:60}), 
(\ref{eq:70}) we use $p_{T}/z$.
We calculate the vacuum FFs $D_{h/j}$  
as a convolution of the KKP \cite{KKP} parton$\to$hadron FFs at 
soft scale $Q_{0}=2$ GeV with the DGLAP parton$\to$parton FFs 
$D_{j/i}^{DGLAP}$
describing the evolution from $Q$ to $Q_{0}$. The latter have been 
computed with the help of PYTHIA \cite{PYTHIA}. 
The medium modified FFs $D_{j/i}^{m}$ have been calculated in a similar way but
inserting between the DGLAP parton$\to$parton FFs and the KKP
parton$\to$hadron
FFs the parton$\to$parton FFs $D_{j/i}^{ind}$ which correspond to 
the induced radiation stage in the QGP.
The $D_{j/i}^{ind}$ have been calculated
from the medium induced gluon spectrum using Landau's 
method \cite{BDMS_RAA} imposing the flavor and momentum conservation
(see \cite{RAA08} for details).
Note that the permutation of the DGLAP and the induced stages 
gives a very small effect \cite{RAA08}.

Since we ignore the azimuthal effects, the averaging of 
the medium modified FFs over the geometrical
variables of the hard parton process and over the impact parameter of 
$pp$ collision is simply reduced to
averaging over the parton path length $L$ in the QGP.
We have performed averaging over $L$ 
for the distribution of hard processes in the impact parameter plane 
obtained with the quark distribution from the MIT bag model
(we assume that the valence quarks and the hard gluons 
radiated by the valence quarks have approximately the same 
distribution in the transverse spacial coordinates).
We obtained that
practically in the full range of the $pp$ impact parameter the distribution
in $L$ is sharply peaked around $L\approx\sqrt{S_{ov}/\pi}$ (here $S_{ov}$
is the overlap area for two colliding bags). It shows that $R_{f}$ 
at the same time gives the typical path length for fast partons.
We found that, as compared to $L=R_{f}$, the $L$-fluctuations reduce the 
medium modification  by only $\sim 10-15$\%.

We treat the collisional energy loss, which is relatively 
small \cite{Z_Ecoll}, as a small perturbation to 
the radiative mechanism, and 
incorporate it simply by renormalizing 
the QGP temperature in calculating the medium modified FFs for
the induced radiation
(see \cite{RAA08} for details).

\subsection{Predictions for $R_{pp}$}
In Fig.~3 we present the results for $R_{pp}$ of charged hadrons 
at $\sqrt{s}=0.2$, $2.76$ and $7$ TeV for $\alpha_{s}^{fr}=0.5$, 
$0.6$ and $0.7$.
To illustrate the sensitivity of the results to $\tau_{0}$
we show the curves 
for $\tau_{0}=0.5$ and $0.8$ fm.
The suppression effect for 
the basic variant with $\tau_{0}=0.5$ fm
turns out to be quite large at $p_{T}\lsim 20$ GeV both for
RHIC and LHC. One can see that for $\tau_{0}=0.8$ fm the reduction
of the suppression is not very significant.
Fig.~3 shows that, as we expected, $R_{pp}$ does not exhibits a 
strong dependence on $\alpha_{s}^{fr}$. 
Although the plasma density is smaller at
$\sqrt{s}=0.2$ TeV,  the suppression effect is approximately similar 
to that at $\sqrt{s}=2.76$ and $7$ TeV. It is due to a steeper slope of the hard
cross sections at $\sqrt{s}=0.2$ TeV. The increase in the suppression
from $\sqrt{s}=2.76$ to $\sqrt{s}=7$ TeV is relatively small.
In the left part of Fig.~4 we show a comparison between $R_{pp}$ at 
$\sqrt{s}=7$ TeV 
for the minimum bias and the UE $dN_{ch}/d\eta$. 
One can see that even the minimum bias $dN_{ch}/d\eta$ gives 
a considerable suppression.
The right part of Fig.~4 
shows variation of $R_{pp}$ between $\sqrt{s}=7$ and $100$ TeV.
One sees that the energy dependence of $R_{pp}$ is weak. 

%
\begin{figure}[ht]
\epsfig{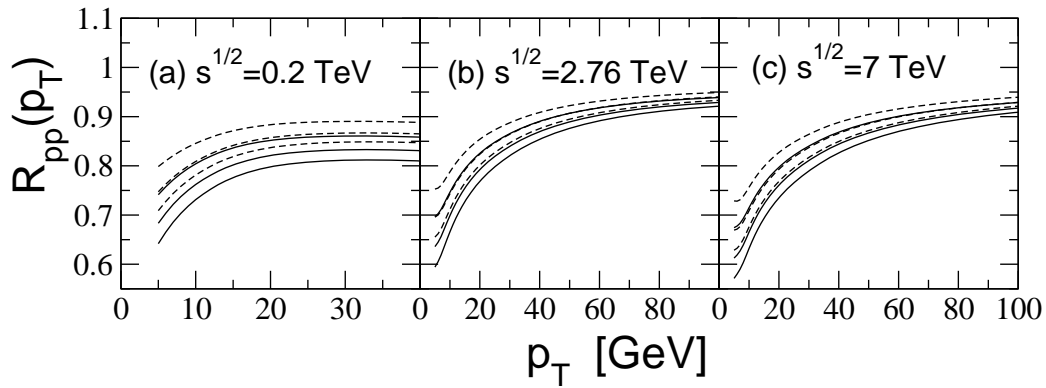}
\caption{$R_{pp}$ of charged hadrons 
at $\sqrt{s}=0.2$ (a), $2.76$ (b), $7$ (c) TeV for (top to bottom)
$\alpha_{s}^{fr}=0.5$, $0.6$ and $0.7$ for $\tau_{0}=0.5$ (solid)
and $0.8$ (dashed) fm.
}
\end{figure}
\begin{figure*}[h]
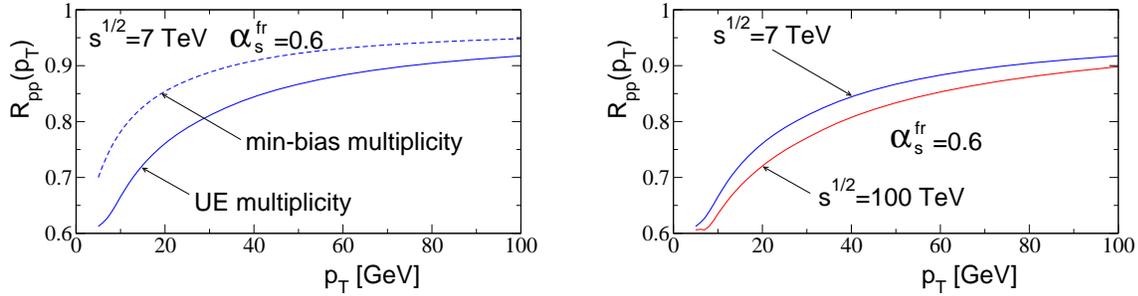

\hspace*{-0.8cm }\epsfig{file=fig2n-talk.eps,height=3.9cm,clip=}
\hspace*{0.8cm }\epsfig{file=fig100tev.eps,height=3.9cm,clip=} 
\begin{minipage}[t]{17.cm}
\caption{
Left:
$R_{pp}$ of charged hadrons at $\sqrt{s}=7$ TeV for the UE (solid line) 
and minimum bias (dashed line) $dN_{ch}/d\eta$. 
Right: $R_{pp}$ of charged hadrons  at $\sqrt{s}=7$ TeV (blue) and $\sqrt{s}=100$ TeV (red)
for UE $dN_{ch}/d\eta$.
}
\end{minipage}
\end{figure*}

To study the sensitivity of $R_{pp}$ to the fireball radius
we also performed the calculations for $R_{f}$
given by (\ref{eq:40}) times $0.7$ and $1.3$. 
We found that in these two cases the medium suppression is 
smaller by $\sim 3$\% and $10$\%, respectively.
The weak dependence on $R_{f}$ is due to a compensation 
between the enhancement of the energy loss caused by increase of the 
fireball size and its suppression due to reduction of the QGP density.
Note that the stability of $R_{pp}$ against variations of $R_{f}$ shows 
that the variation 
of the plasma density in the transverse coordinates should  not be 
very important.
Indeed, the gluon spectrum is dominated by $N=1$ rescattering term which
is a linear functional of the 
plasma density profile along the fast parton trajectory. 
Therefore the energy loss for a more realistic plasma density 
(with a higher density 
in the central region)  can be roughly approximated
by a linear superposition of that  for the step 
density distributions with different $R_{f}$. And it should 
not change strongly $R_{pp}$ as compared to our calculations.

Fig.~3 shows the results for the typical UE 
multiplicity density.
An accurate accounting for the fluctuations
of the UE $dN_{ch}/d\eta$ is impossible since it should be done 
on the event-by-even basis, and
requires detailed information about dynamics of the 
UEs.
To understand how  the
event-by-event fluctuations of the UE $dN_{ch}/d\eta$ may change our results,
we evaluated $R_{pp}$ assuming 
that the distribution in the UE $dN_{ch}/d\eta$ is the same 
at each impact parameter and jet production point. We used 
the distribution in $dN_{ch}/d\eta$ from CMS
\cite{CMS_UE_Nch} measured at $\sqrt{s}=0.9$ and $7$ TeV. 
It satisfies approximately KNO scaling
similar to that in minimum bias events \cite{Dumitru_KNO}.
For this reason one can expect that it can be used for RHIC conditions
as well. We observed that the fluctuating  $dN_{ch}/d\eta$
suppresses $(1-R_{pp})$ by only $\sim 5-6$\% 
both for RHIC and LHC energies. 
This says that our approximation without the event-by-event
fluctuations of the QGP parameters
should be good.

\subsection{Effect of $R_{pp}$ on $R_{AA}$}
To illustrate the effect of the mini-QGP
in $pp$ collisions on $R_{AA}$ in Fig.~5
we compare our results for $R_{AA}$
with the data for $\pi^{0}$-mesons in central $Au+Au$ 
collisions at $\sqrt{s}=0.2$ TeV (a) from PHENIX \cite{PHENIX_RAA_pi},
and with the data for charged hadrons in central $Pb+Pb$ collisions
at $\sqrt{s}=2.76$ TeV (b,c)
from ALICE \cite{ALICE_RAAch} and CMS \cite{CMS_RAAch}. 
%
\begin{figure} [hb]
\epsfig{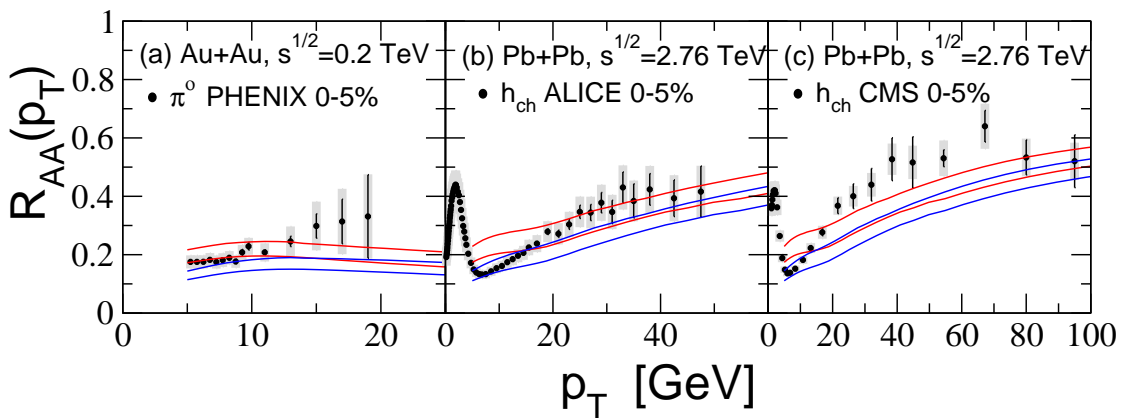}
\caption{(a) $R_{AA}$ of $\pi^{0}$ for 0-5\% central $Au+Au$ collisions
at $\sqrt{s}=0.2$ TeV from our calculations 
for (top to bottom) $\alpha_{s}^{fr}=0.5$ and $0.6$
with (red) and without (blue) $1/R_{pp}$ factor in (\ref{eq:21}).
(b,c) $R_{AA}$ for charged hadrons for 0-5\% central $Pb+Pb$ collisions
at $\sqrt{s}=2.76$ TeV from our calculations 
for (top to bottom) $\alpha_{s}^{fr}=0.4$ and $0.5$
with (red) and without (blue) $1/R_{pp}$ factor in (\ref{eq:21}).
The red curves are obtained with the factor $1/R_{pp}$ 
calculated with $\alpha_{s}^{fr}=0.6$.
Data points are from  PHENIX \cite{PHENIX_RAA_pi} (a),
ALICE \cite{ALICE_RAAch} (b) and CMS \cite{CMS_RAAch} (c).
Systematic experimental errors are shown as shaded areas. 
}
\end{figure}
%
\begin{figure}[h]
\epsfig{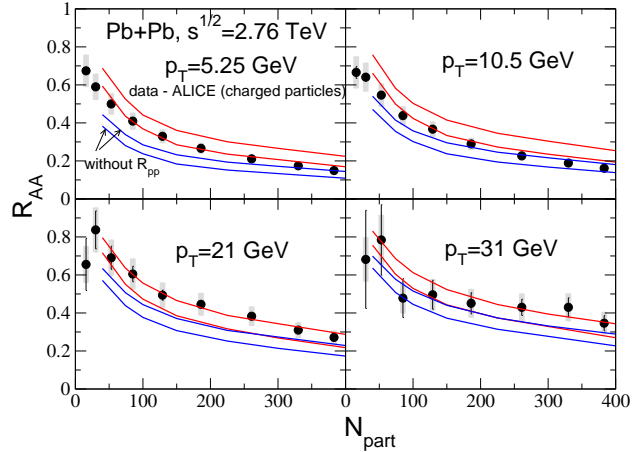}
\caption{$R_{AA}$  of charged particles vs $N_{part}$ for $Pb+Pb$ at 
$\sqrt{s}=2.76$ TeV
with (red) and without (blue) $R_{pp}$, for
(top to bottom) $\alpha_{s}^{fr}=0.4$ and $0.5$
for $\sqrt{s}=2.76$ TeV,
$R_{pp}$ is calculated at $\alpha_{s}^{fr}=0.6$.
Data points are from ALICE \cite{ALICE_RAA_Nch}.}
\end{figure}
%
%
\begin{figure}
\epsfig{file=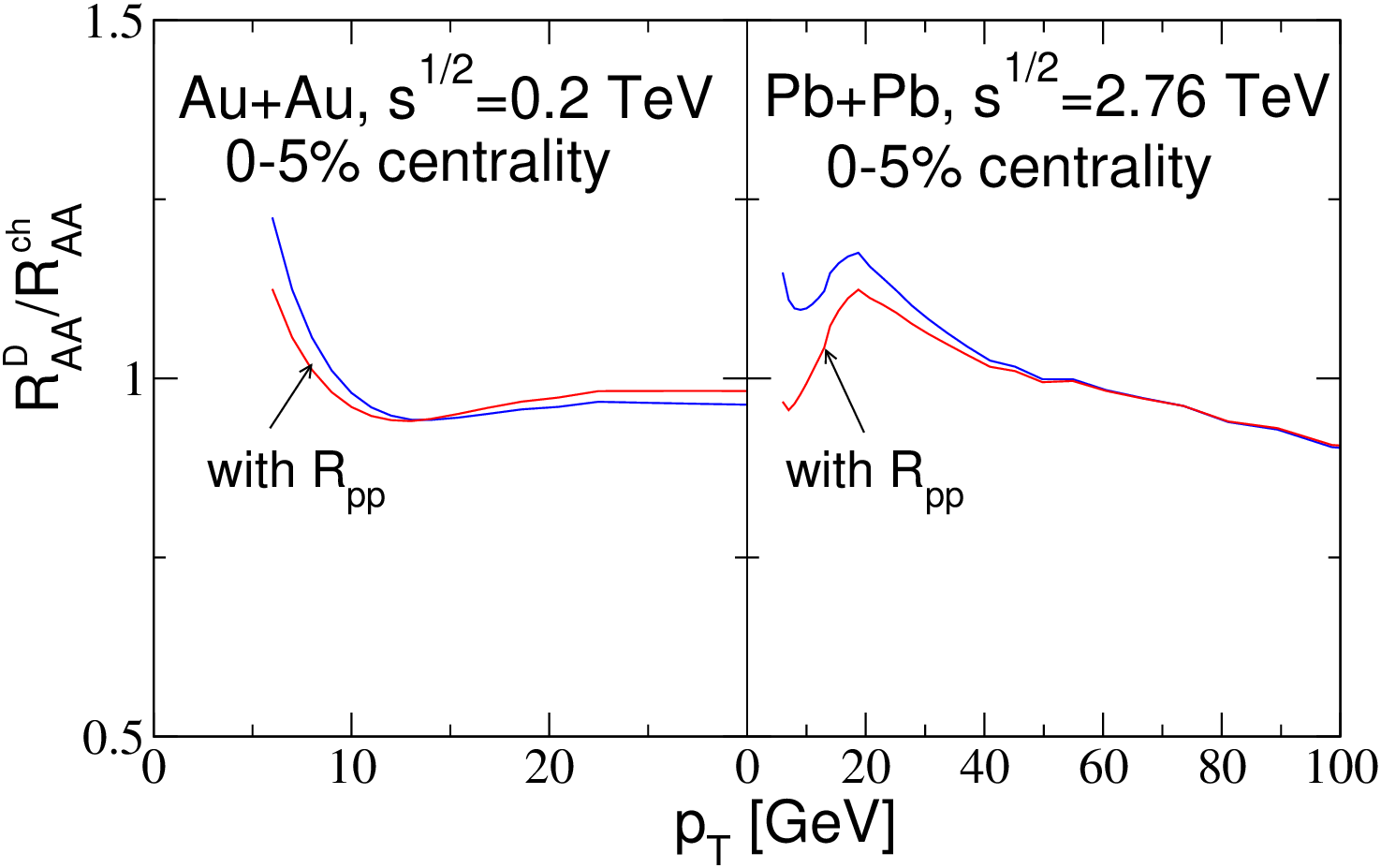,height=5cm,clip=}
\caption{Effect of $R_{pp}$ due to mini-QGP on ratio $R_{AA}$ for $D$-mesons to $R_{AA}$ for light charged 
hadrons. $\alpha_{s}^{fr}=0.6$ for $\sqrt{s}=0.2$ TeV and $\alpha_{s}^{fr}=0.5$
for $\sqrt{s}=2.76$ TeV,
$R_{pp}$ for light and heavy flavors is calculated at $\alpha_{s}^{fr}=0.6$.}
\end{figure}
\begin{figure}
\epsfig{file=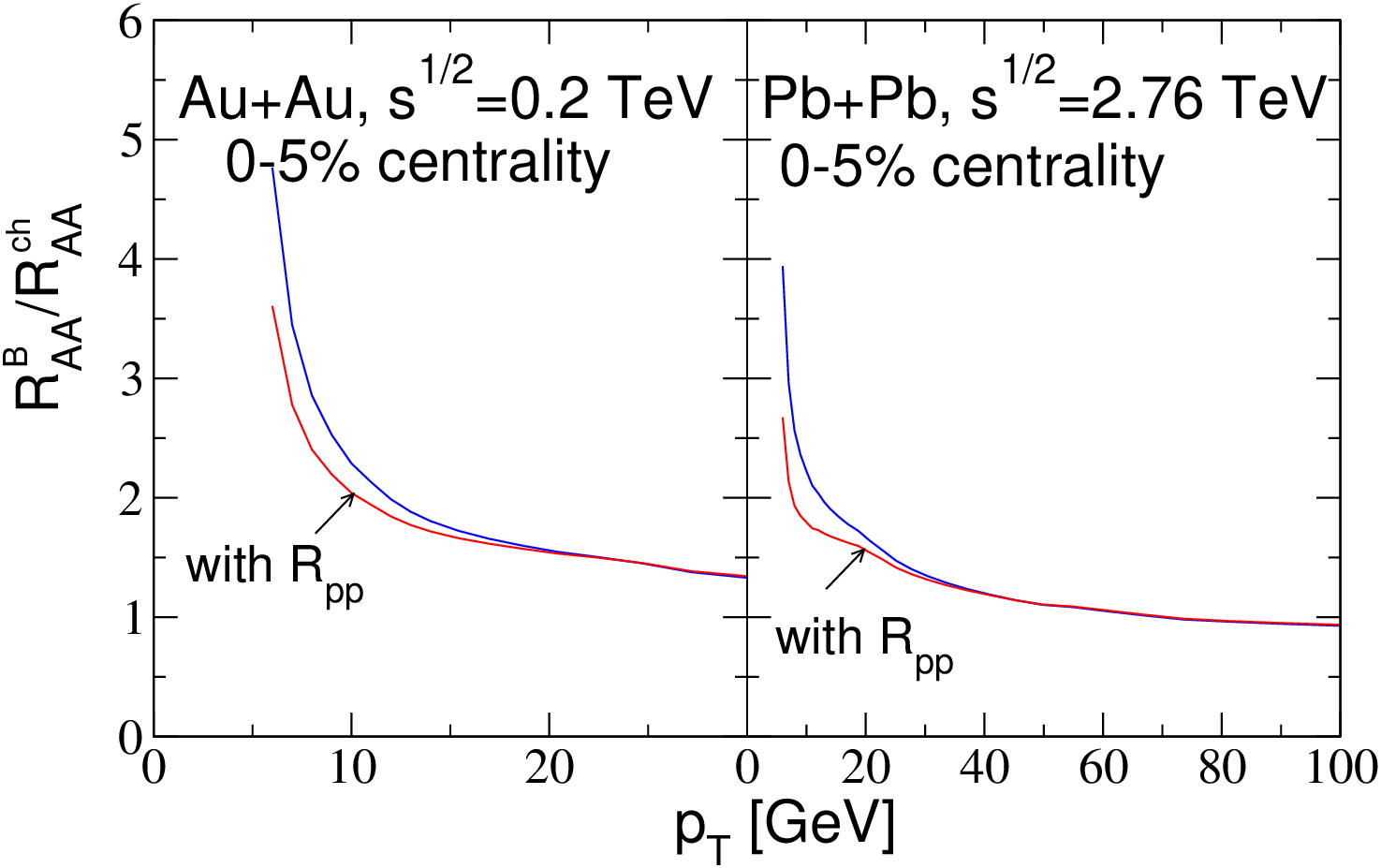,height=5cm,clip=}
\caption{Effect of $R_{pp}$ due to mini-QGP on ratio $R_{AA}$ for $B$-mesons to $R_{AA}$ for light charged 
hadrons. $\alpha_{s}^{fr}=0.6$ for $\sqrt{s}=0.2$ TeV 
and $\alpha_{s}^{fr}=0.5$
for $\sqrt{s}=2.76$ TeV,
$R_{pp}$ for light and heavy flavors is calculated at $\alpha_{s}^{fr}=0.6$.}
\end{figure}
%
\begin{figure}[hb]
\epsfig{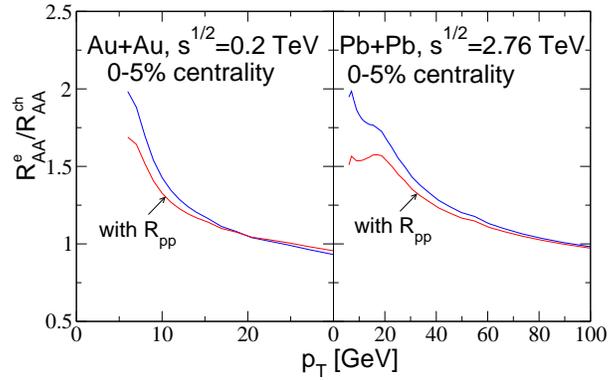}
\caption{Effect of $R_{pp}$ due to mini-QGP on ratio $R_{AA}$ for non-photonic
electrons to $R_{AA}$ for light charged 
hadrons. $\alpha_{s}^{fr}=0.6$ for $\sqrt{s}=0.2$ TeV 
and $\alpha_{s}^{fr}=0.5$
for $\sqrt{s}=2.76$ TeV,
$R_{pp}$ for light and heavy flavors is calculated at $\alpha_{s}^{fr}=0.6$.}
\end{figure}
%
\begin{figure}[!hb]
\epsfig{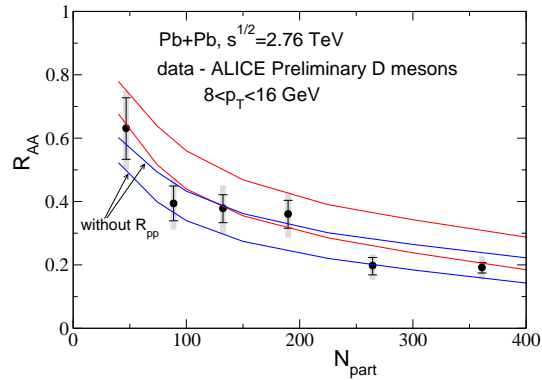}
\caption{$R_{AA}$  of $D$-mesons vs $N_{part}$ for $Pb+Pb$ at $\sqrt{s}=2.76$ TeV
with (red) and without (blue) $R_{pp}$,
for (top to bottom)  $\alpha_{s}^{fr}=0.4$ 
and $\alpha_{s}^{fr}=0.5$,
$R_{pp}$ is calculated at $\alpha_{s}^{fr}=0.6$.
Data points are from ALICE \cite{ALICE_D_Nch}.}
\end{figure}
We show the predictions for $R_{AA}$ defined by (\ref{eq:21})
with (red) the $1/R_{pp}$ factor,
and for $R_{AA}^{st}$ without (blue) this factor.
We use the $R_{pp}$ for $\alpha_{s}^{fr}=0.6$.
We calculated $R_{AA}^{st}$ for
$\alpha_{s}^{fr}=0.5$ and $0.6$ at $\sqrt{s}=0.2$ TeV, 
and for $\alpha_{s}^{fr}=0.4$ and $0.5$
at $\sqrt{s}=2.76$ TeV. Because these values give  better agreement 
with the data. 
We accounted for the nuclear modification of the PDFs
with the EKS98 correction \cite{EKS98}.
As in \cite{RAA13}, we take $T_{0}= 320$ MeV for central
$Au+Au$ collisions at $\sqrt{s}=0.2$ TeV, and
$T_{0}= 420$ MeV for central
$Pb+Pb$ collisions at $\sqrt{s}=2.76$ TeV
obtained from hadron multiplicity pseudorapidity density $dN_{ch}/d\eta$ 
from RHIC \cite{STAR_Nch} and LHC \cite{CMS_Nch,ALICE_Nch}.
At $p_{T}\sim 10$ GeV for RHIC the agreement 
of the theoretical $R_{AA}$
(with the  $1/R_{pp}$ factor) with the data  
is somewhat better for $\alpha_{s}^{fr}=0.6$, and for LHC 
the value $\alpha_{s}^{fr}=0.5$ seems to be preferred by the data.
The agreement in the $p_{T}$-dependence
of $R_{AA}$ is not perfect (especially for LHC).
The theory somewhat
underestimates the slope of the data. 
It seems that the regions of large $p_{T}$ support 
$\alpha_{s}^{fr}=0.5$ and $0.4$ for RHIC and LHC, respectively.
The inclusion of $R_{pp}$ even reduces  a little  the slope of $R_{AA}$.
However, it does not seem to be very dramatic
since the theoretical uncertainties 
may be significant.

Fig.~5 shows that the effect of $R_{pp}$ on $R_{AA}$ 
in central $AA$ collisions can approximately be imitated by 
a simple reduction of $\alpha_{s}^{fr}$.
However, it is clear that $R_{pp}$ may be 
important for the azimuthal effects and the centrality dependence of 
$R_{AA}$ since 
in the scenario with the mini-QGP formation 
in $pp$ collisions the values of $\alpha_{s}^{fr}$ become bigger.
The effect of $R_{pp}$ on the centrality dependence of $R_{AA}$ is shown
Fig.~6. 
$R_{pp}$ can  also affect the flavor dependence of 
$R_{AA}$ since the suppression effect for heavy quarks 
in $pp$ collisions is smaller. 
It is illustrated in Figs.~7--9 for the $p_{T}$-dependence of the ratio
of the $R_{AA}$ for heavy and light flavors. 
One sees that at $p_{T}\lsim 10$ GeV $R_{pp}$ reduces the difference between
the nuclear suppression of the spectra for heavy and light flavors. 
In Fig.~10 we show the 
effect of $R_{pp}$  on the centrality dependence of $R_{AA}$ for $D$-mesons.
One can see that $R_{pp}$ may improve somewhat agreement with the data.

\subsection{Jet quenching in $pA$ collisions}
In the scenario with the QGP production in $pp$ collisions
the correct formula for $R_{pA}$ reads
 $R_{pA}=R_{pA}^{st}/R_{pp}$.
Evidently, the sizes and the initial temperatures of the plasma fireballs in 
$pp$ and $pA$ collisions should not differ strongly. 
For this reason for $R_{pA}$ the uncertainties related to variation of
$\alpha_{s}$ (or the temperature dependence of the QGP
 density and the Debye mass) are smaller than for $R_{AA}$.
The ALICE data \cite{ALICE_RpPb} show a small deviation from unity 
of $R_{pPb}$ at $\sqrt{s}=5.02$ TeV at $p_{T}\gsim 10$ GeV, where the Cronin effect should be weak. 
In the scenario with the QGP formation 
this is possible only if the magnitudes of the medium suppression
in $pp$ and $pPb$ collisions are close to each other.
Unfortunately, presently the UE multiplicity in 
$pPb$ collisions is unknown. 
But it is clear that it cannot be smaller
than the minimum bias multiplicity density 
$dN_{ch}^{mb}/d\eta=16.81\pm 0.71$ \cite{ALICE_dnch_pPb}.
In order to understand the acceptable range of the
UE multiplicity density in $pPb$ collisions in the scenario with the mini-QGP
formation we calculated $R_{pPb}$ for 
$dN_{ch}/d\eta=K_{ue}dN_{ch}^{mb}/d\eta$ for 
$K_{ue}=1$, $1.25$, and $1.5$.

In our calculations as a basic choice we use the parametrization of  
$R_{f}(pPb)$ vs the multiplicity given in \cite{RPP}
obtained from the results of simulation of the $pPb$ collisions  
performed in \cite{glasma_pp}    
within the IP-Glasma model \cite{IPG12}.
Ref. \cite{IPG12} gives $R_{f}(pPb)$ that is close to $R_{f}(pp)$
where $R_{f}(pp)\propto(dN_{g}/dy)^{1/3}$, but $R_{f}(pPb)$ flattens 
at higher values of the gluon density. 
Using formula (\ref{eq:30}), we obtained
for $K_{ue}=[1,1.25,1.5]$
\beq
R_{f}(pPb)\approx[1.63,1.88,1.98]\,\,\mbox{fm}\,,
\label{eq:80}
\eeq
\beq
T_{0}(pPb)\approx[222,229,235]\,\,\mbox{MeV}\,.
\label{eq:90}
\eeq

Fig.~11 shows comparison of our results with the data on  
$R_{pPb}$ at $\sqrt{s}=5.02$ TeV
from ALICE \cite{ALICE_RpPb}.
To illustrate the sensitivity to $R_{f}(pPb)$
we also present the results for $R_{f}(pPb)$ $1.2$ and $1.4$ times
greater.
We show the curves with (red) and without
(blue) the $1/R_{pp}$ factor.
As for $AA$ case we account for the nuclear modification
of the PDFs with the EKS98 correction \cite{EKS98}. It
gives a small deviation of $R_{pPb}$ from unity even without
parton energy loss.
The results for $R_{pp}$ are also shown (green).
All the curves are obtained with $\alpha_{fr}=0.6$.
However, our predictions for $R_{pPb}$ (with the $1/R_{pp}$ factor)
are quite stable against variation of $\alpha_{s}^{fr}$ since
the medium effects are very similar for $pp$ and $pPb$ collisions.

\begin{figure} [!hb]
\epsfig{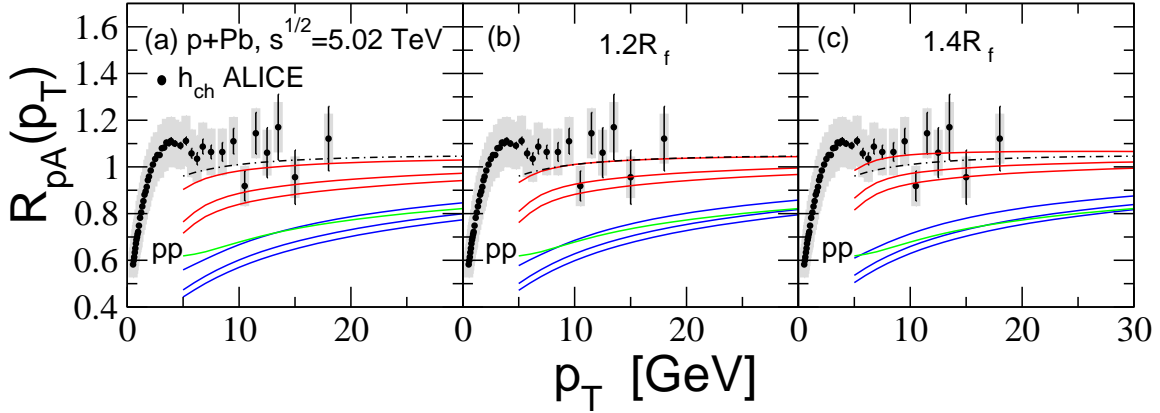}
\caption{(a) $R_{pPb}$ for charged hadrons at 
$\sqrt{s}=5.02$ TeV from our calculations for $\alpha_{s}^{fr}=0.6$
with (red) and without (blue) the $1/R_{pp}$ factor
for (top to bottom)  $K_{ue}=1$, $1.25$ and $1.5$ 
for the $R_{f}(pPb)$ from (\ref{eq:80}).  
(b,c) same as (a) but for $R_{f}(pPb)$ times $1.2$ and $1.4$.
The green line shows $R_{pp}$. The dot-dashed line
shows $R_{pPb}$  due to the EKS98 correction
\cite{EKS98} to the nucleus PDFs.
Data points are from ALICE \cite{ALICE_RpPb}. 
}
\end{figure}
%
Fig.~11 shows that at $p_{T}\gsim 10$ GeV, where the 
Cronin effect should be small, our predictions (with $1/R_{pp}$ factor) 
obtained with $K_{ue}=1$ agree qualitatively with the data. The agreement 
becomes better for larger $R_{f}(Pb)$. But just as for $R_{pp}$ the variation
of $R_{pPb}$ with the fireball size is relatively weak.
The curves for the higher UE multiplicities ($K_{ue}=1.25$ and $1.5$) 
lie below the data. 
Thus we see that the data from ALICE \cite{ALICE_RpPb}
may be consistent with the formation of the QGP in $pp$ and $pPb$ collisions
if the UE multiplicity is close to the minimum bias one.
This condition may be weakened if the size of the fireball in $pPb$ 
collisions is considerably bigger than predicted in \cite{glasma_pp}.
But the physical picture may change if we take into account the meson-baryon
Fock component in the proton.
Indeed, in $pA$ collisions the final-state interaction may be smaller due to
meson-baryon Fock component in the proton.
The weight of the $MB$-component may be as large as $\sim 40$\%
\cite{Speth}. 
Contrary to $pp$ case in $pA$ collisions practically in all events 
meson should produce its own fireball. 
It means that in $\sim 40$\% events an asymmetric
two-fireball configuration may be produced (as illustrated in the left 
part of Fig.~ 12). 
Since jet may propagate without interaction with
one of the fireball (typically it is the meson fireball as shown in the right 
part of Fig.~12), 
the final-state interaction should be weaker
than for a symmetric fireball (for same $dN_{ch}/d\eta$).
Note that the two-fireball state naturally generates the azimuthal flow
for the soft particles as well.
\begin{center}
\begin{figure} [h]
\epsfig{file=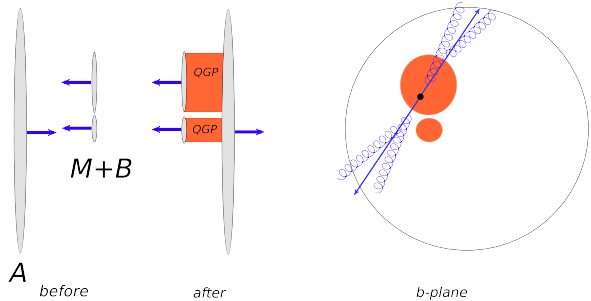,height=4cm,clip=}
\caption{A cartoon of the production of a two-fireball state in $pA$ collisions from
the meson-baryon Fock component of the proton (a); A carton of the 
jet quenching for the two-fireball state (b)}.
\end{figure}
\end{center}

\section{Medium modification of photon-tagged and inclusive jets in 
high-multiplicity proton-proton collisions}
For a direct observation of the medium effects in $pp$ collisions  
one can use measurement of the jet FF
in $\gamma+$jet events for different UE multiplicities.
To understand the prospects 
of this method 
we evaluate the medium modification
of the $\gamma$-tagged FF at $\sqrt{s}=7$ TeV 
at $y=0$.
The values of the $R_{f}$ and $T_{0}$ for 
different values of $dN_{ch}/d\eta$ obtained using (\ref{eq:30}) 
are given in Table I.
For $dN_{ch}/d\eta\gsim 40$ we obtain $T_{0}$ which is about that for 
central $Au+Au$ collisions at RHIC.
\begin{table}[h]
\caption{$R_{f}$ and $T_{0}$ for different $dN_{ch}/d\eta$.}
\begin{tabular}{|c|c|c|c|c|c|} 
\hline
$dN_{ch}/d\eta$ & 3 & 6 & 20 & 40 & 60 \\
\hline
$R_{f}$ (fm) & 1.046 & 1.27 & 1.538 & 1.538  & 1.538 \\
\hline
$T_{0}$ (MeV) & 177 & 196 & 258 & 325 & 372 \\
\hline
\end{tabular}
\end{table}

In $\gamma$+jet events
the energy of the hard parton, $E_{T}$, in the direction opposite 
to the tagged photon is smeared around the photon energy, $E_{T}^{\gamma}$.
But using the results of the NLO calculations
\cite{Wang_NLO2} one can show that at $E_{T}^{\gamma}\gsim 25$ GeV
and $z\lsim 0.9$ the smearing can be safely neglected (for details, see
\cite{Z_phot}).
To be conservative we present results for $z<0.8$, where the effect of
smearing is practically negligible and one can set $E_{T}=E_{T}^{\gamma}$.
Then, as in \cite{W12}, 
we can write the $\gamma$-tagged FF as a 
function of the UE multiplicity density $dN_{ch}/d\eta$ (for
clarity we denote it by $N$) as
\beq
D_{h}(z,E_{T}^{\gamma},N)\!=\!\big\langle\!\big\langle
\!\sum_{i} r_{i}(E_{T}^{\gamma})D_{h/i}^{m}(z,E_{T}^{\gamma},N)
\!\big\rangle\!\big\rangle,
\label{eq:100}
\eeq
where, as in (\ref{eq:70}),
$D_{h/i}^{m}$ is the medium modified FF for $i\to h$ process, and 
$r_{i}$ is the fraction of the $\gamma+i$ parton state in the 
$\gamma+$jet events, $\langle\!\langle...\rangle\!\rangle$ means averaging
over the transverse geometrical variables of $pp$ collision 
and jet production, which includes averaging over the fast parton path length
$L$ in the QGP.
Just as for $R_{pp}$ we have performed averaging over $L$ 
using the distribution of hard processes in the impact parameter plane 
obtained with the quark distribution from the MIT bag model.
As compared to $L=R_{f}$ the $L$-fluctuations
reduce the medium modification by $\sim 10-15$\%.
In Fig.~13 we present the results for the medium modification factor
(for charged hadrons)
\beq
I_{pp}(z,E_{T},N)=D_{h}(z,E_{T},N)/D^{vac}_{h}(z,E_{T})\,
\label{eq:110}
\eeq
for the $\gamma$-tagged (upper panels)  jets
for  $E_{T}=[25, 50, 100]$ GeV at $\sqrt{s}=7$ TeV.
For comparison we also show the results for inclusive (lower panels) jets.  
The smearing effect is irrelevant to inclusive jets and we show
the results for the whole range of $z$.
For illustration of the difference between $pp$ and $AA$ collisions 
we also present the curves for 
$\sqrt{s}=2.76$ TeV  for $L=5$ fm and $T_{0}=420$ MeV that can be regarded 
as reasonable values for $Pb+Pb$ collisions (we used  $\alpha_{s}^{fr}=0.4$, 
which is favored by the data on $R_{AA}(p_{T})$ at $p_{T}\gsim 20$ GeV).
Fig.~13 shows that there is a considerable quenching effect for 
$dN_{ch/d\eta}\gsim 20$. 
Note that the observed strong quenching of inclusive jets is 
qualitatively supported by the preliminary data from 
ALICE \cite{ALICE_jet_UE} that indicate
that for the high multiplicity UEs jets undergo a softer fragmentation.
\begin{figure}
\vspace{.7cm}
\epsfig{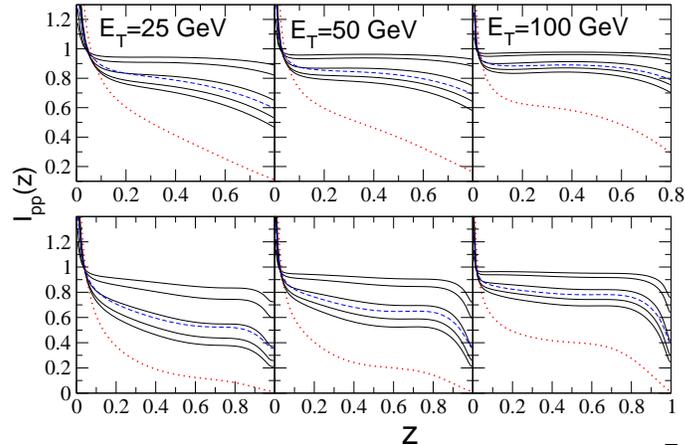}
\vspace{-0.5cm}
\caption{$I_{pp}$ for $\gamma$-tagged
(upper panels) and inclusive (lower panels)  jet FFs  
at $\sqrt{s}=7$ TeV for $dN_{ch}/d\eta=[3, 6, 20, 40, 60]$ 
(solid line). The order (top to 
bottom) of the curves
at large $z$ corresponds to increasing values of
$dN_{ch}/d\eta$.
The dashed blue line shows ratio of the FFs for
$dN_{ch}/d\eta=40$ and $3$. The red dotted line shows the medium
modification factor at $\sqrt{s}=2.76$ TeV for the QGP 
with $T_{0}=420$ MeV and $L=5$ fm for $\alpha_{s}^{fr}=0.4$.
}
\end{figure}

%
Since the vacuum FFs are unobservable,
in practice, to observe the medium effect 
one should simply compare the FFs
for different multiplicities. 
In Fig.~13 we show the ratio of the FFs for $N=40$ and $N=3$
(for inclusive jets this ratio cannot be measured, and we show it just 
to illustrate the difference in magnitudes of the effect for 
$\gamma$-tagged and inclusive jets).
As for $R_{pp}$ we have investigated the sensitivity of our results 
to variation of $R_{f}$, and found that $I_{pp}$ 
is quite stable against  variation of $R_{f}$.

\section{Summary}
Assuming that a mini-QGP fireball may be created in
$pp$ collisions, we have evaluated 
the medium modification of high-$p_{T}$ particle
spectra for light and heavy flavors and medium modification factors
for the $\gamma$-triggered and inclusive jet FFs.
For $p_{T}\sim 10$ GeV we obtained
$R_{pp}\sim [0.7-0.8,\,0.65-0.75,\,0.6-0.7]$ 
at $\sqrt{s}=[0.2, 2.76,7]$ TeV.
We have studied the effect of $R_{pp}$ on the
theoretical predictions for
the nuclear modification factor $R_{AA}$ in $AA$ collisions at 
RHIC and LHC energies. We found that $R_{pp}$ does not change dramatically 
the description of the data on  $R_{AA}$ for light hadrons
in central $AA$ collisions, and its effect may be imitated by some 
renormalization of $\alpha_{s}$.
But inclusion of $R_{pp}$  changes the centrality dependence of $R_{AA}$.
Also, $R_{pp}$ weakens the flavor dependence of $R_{AA}$.

Our results show that the ALICE data \cite{ALICE_RpPb}  
on $R_{pPb}$ may be consistent
with the scenario with the QGP formation if 
in $pPb$ collisions the UE multiplicity is close to the minimum bias one. 
But this condition may be weakened due to presence in the proton wave function
of the meson-baryon Fock component. 
We leave analysis of its effect for future work.  

We demonstrated that in $pp$ collisions
with UE multiplicity density $dN_{ch}/d\eta\sim 20-40$
the mini-QGP  can suppress
the $\gamma$-triggered FF at $E_{T}\sim 25-100$ GeV and $z\sim 0.5-0.8$
by $\sim 10-40$\%, and for inclusive jets the effect is even stronger.
%

\begin{acknowledgments}
I would like to thank P. Arnold for the invitation to give
this talk at this XIth Quark Confinement and the Hadron Spectrum 
International Conference.
This work is supported 
in part by the 
grant RFBR
12-02-00063-a and the program
SS-6501.2010.2.
\end{acknowledgments}

\end{document}